\documentclass[prl,aps,showpacs,twocolumn,preprintnumbers]{revtex4}
\usepackage{amsfonts}
\usepackage{amsmath}
\usepackage{amssymb}

\usepackage{dcolumn}

\usepackage{amsfonts}
\usepackage{amsmath}
\usepackage{amssymb}

\def\Z{\mathbb Z}

\def\R{\mathbb R}

\begin{document}

\title{ \textbf{Hidden supersymmetry in  quantum bosonic systems}}
\author{\textsf{Francisco Correa and Mikhail S. Plyushchay}\\
\vskip 0.1cm
 {\textit{Departamento de F\'{\i}sica, Universidad de
Santiago de Chile, Casilla 307, Santiago 2, Chile\\\vskip 0.15cm
E-mails: francisco.correa@usach.cl, mplyushc@lauca.usach.cl}} }

\begin{abstract}
We show that some simple well studied quantum mechanical systems
without fermion (spin) degrees of freedom display, surprisingly,
a hidden supersymmetry. The list includes the bound state
Aharonov-Bohm, the Dirac delta and the P\"oschl-Teller potential
problems, in which the unbroken and broken $N=2$ supersymmetry of
linear and nonlinear (polynomial) forms is revealed.
\end{abstract}

 \maketitle


\emph{I. Introduction.}  ---Supersymmetry (SUSY),   as a symmetry
between bosons and fermions, was originally introduced in search of
a nontrivial unification of space-time and internal symmetries in
relativistic quantum field theory \cite{SUSY}. To explain its no (so
far) experimental evidence in nature, supersymmetric quantum
mechanics was invented by Witten as a toy model to investigate a
SUSY breaking in field theory \cite{Witten}. Subsequently
supersymmetric quantum mechanics was transformed into independent
line of research, which stimulated new approaches to other branches
of physics including atomic, nuclear, condensed matter and
statistical physics \cite{CKS}.

The low dimensional physics possesses some peculiar features among
which are a remarkable equivalence between fermions and bosons in 2D
field theories \cite{Stone}, and a boson-fermion, or more generally,
a boson-anyon transmutation   based on the Aharonov-Bohm  (AB)
effect in planar systems \cite{Anyons}. These peculiarities indicate
that SUSY could be present in a hidden form in some bosonic systems.
It really was observed in quantum mechanical models with a
\emph{nonlocal} Hamiltonian depending on a reflection (parity, or
exchange) operator \cite{SUSYbos}, and in a related parabosonic
system \cite{ParaP}. On the other hand, hidden SUSY of a nonlinear
(polynomial) form \cite{AIS,ParaP} was recently found
\cite{SUSYconf} in a conformal mechanics model \cite{confmec}
described by a local Hamiltonian. However, it appears there as a
\emph{fictitious} symmetry due to a boundary conditions breaking
associated with its odd generators.

In this paper we show that some simple well studied quantum
mechanical systems without fermion (spin) degrees of freedom
display, surprisingly, a \emph{true} hidden SUSY. The list of {\it
bosonic} systems with local Hamiltonians includes the bound state
Aharonov-Bohm, the Dirac delta and the P\"oschl-Teller (PT)
potential problems, in which we reveal the unbroken and broken
$N=2$ SUSY of linear and nonlinear (polynomial) forms.


\emph{II. Hidden SUSY in bound state AB effect.} ---Let us first
consider a free particle on a circle of unit radius, given by the
Hamiltonian ($\hbar=2m=1$)
\begin{equation}\label{Hc0}
    H=-\,\frac{d^2}{d\varphi^2}\,.
\end{equation}
The
$2\pi$-periodic eigenfunctions of the angular momentum
$p_\varphi=-i\frac{d}{d\varphi}$,
\begin{equation}
    \psi _{l}(\varphi )=e^{il\varphi },
    \qquad p_{\varphi }\psi _{l}=l\psi_{l},
    \label{mom}
\end{equation}
$l=0,\pm 1,\ldots,$ provide us with a complete basis for the Hilbert
space of states, and solve the spectral problem, $H\psi
_{l}=E_{l}\psi _{l},$ $E_{l}=l^{2}$. All the energy levels are
positive and doubly degenerate except the level $E_0=0$ of the
singlet ground state. Such spectral properties are typical for a
quantum mechanical system having the unbroken $N=2$ SUSY. Though
system (\ref{Hc0}) has no fermion degrees of freedom,  a complete
supersymmetric structure can be revealed in it by identifying a
reflection, $R\psi (\varphi )=\psi (-\varphi ),$ as a grading
operator.
 Indeed, it is a self-adjoint integral of motion,
$R=R^\dagger$, $R^2=1$, $[H,R]=0$, which anticommutes with angular
momentum. Hence the self-adjoint operators
\begin{equation}
    Q_{1}=p_{\varphi },\qquad Q_{2}=iRQ_{1},
    \label{Qc}
\end{equation}
generating  an ordinary  $N=2$ superalgebra,
\begin{equation}
    \{Q_{a},Q_{b}\}=2\delta _{ab}H,\qquad \lbrack H,Q_{a}]=0,
    \label{Susya}
\end{equation}
are identified as the supercharges. The energy eigenstates $\psi
_{l}^{+}=\cos l\varphi $ and $\psi _{l}^{-}=\sin l\varphi $,
satisfying the relations $R\psi _{l}^{\pm }=\pm \psi _{l}^{\pm }$,
play here a role of the bosonic- and fermionic-like states. The
peculiarity of the described  SUSY of this simple system with pure
bosonic local  Hamiltonian is hidden in the nonlocal nature of the
grading operator and of one of its two supercharges, namely, $Q_2$.

Consider now a charged particle on a unit circle $x^2+y^2=1$ placed
in the $z=0$ plane  and pierced by a magnetic field of a flux line,
$B_z=\epsilon_{ij}\partial_iA_j=\Phi\,\delta^2(x,y)$, where
$A_i=-\frac{\Phi}{2\pi r^2}\,\epsilon_{ij}r_j $ is a planar vector
potential.
 The Hamiltonian of this
system,
\begin{equation}  \label{HABc}
    H_\alpha=(p_i-\frac{e}{c}A_i)^2=(p_\varphi+\alpha)^2,
\end{equation}
$\alpha= -\,\frac{e}{2\pi c}\Phi$, corresponds to the bound state AB
effect \cite{AB,PesT}. States (\ref{mom}) are  the eigenstates of
Hamiltonian (\ref{HABc}) with eigenvalues $E_{l}=(l+\alpha )^{2}$.
The shifted angular momentum operator $p_\varphi +\alpha$ and
Hamiltonian (\ref{HABc}) can be related to the same operators of the
free  particle ($\alpha=0$) by a transformation
\begin{equation}  \label{AB-0}
    O_\alpha=U_{-\alpha}(\varphi)O_0U_\alpha(\varphi),\qquad
    U_\alpha(\varphi)=e^{i\alpha\varphi}.
\end{equation}
In a generic case relation (\ref{AB-0}) has, however, a formal
character since the unitary-like operator $U_\alpha(\varphi)$ takes
out the states (\ref{mom}) from the Hilbert space of $2\pi$-periodic
wave functions.

When the parameter $\alpha $ takes an integer value $\alpha =n$,
$n\in \mathbb{Z}$, the spectrum reveals the structure of the
unbroken $N=2$ SUSY: the states $\psi _{l}$ and $\psi _{l^{\prime
}}$ with $l^{\prime }=-(l+2n)$, $l\neq -n$, have the same energy,
while the state $\psi _{-n}$ is a singlet ground state of  zero
energy. The twisted reflection operator $R_n=e^{-2in\varphi}R $
plays here the role of the grading operator, allowing us to realize
the supercharge operators of the hidden $N=2$ SUSY in the form
similar to (\ref{Qc}) but with $p_\varphi$ changed for  $p_\varphi
+n$. The supercharges annihilate the singlet bosonic-like state
$\psi_{-n}$, $R_n\psi_{-n}=\psi_{-n}$. Since the $U_{n}(\varphi )$
is a well defined operator in the Hilbert space of $2\pi$-periodic
wave functions, system (\ref{HABc}) turns out to be unitary
equivalent to the free system.

The spectrum is also degenerate in the nontrivial case of the AB
effect characterized by the half-integer values of the parameter
$\alpha$. At $\alpha=n+\frac{1}{2}\equiv j$, there is no singlet
state of zero energy in the spectrum, and all the energy levels are
doubly degenerate, $E_{l}=E_{-(l+2j)}$. This picture corresponds to
the broken $N=2$ SUSY. Though transformation (\ref{AB-0}) in this
case is of a formal character, it produces a well defined grading
operator being a twisted reflection operator $ R_{j}=e^{-i2j\varphi
}R,$ $R_{j}\psi_{j,l}^\pm=\pm \psi_{j,l}^\pm$, where
$\psi_{j,l}^\pm=\psi _{l}\pm
    \psi_{-(l+2j)}$. Having in mind relation (\ref{AB-0}), we find that the Hamiltonian
$H_{j}$ and the supercharges $ Q_{j,1}=p_{\varphi }+j,$
$Q_{j,2}=iR_{j}Q_{j,1}$ are the even and odd operators with
respect to $R_j$, and that they generate the superalgebra of the
form (\ref{Susya}). In both cases $\alpha=n,j$ the local
supercharge $Q_1$ is diagonal on the states (\ref{mom}).

Note here that the special ``magic"  of half fluxons and resulting
degeneracy was discussed in the context of the AB and Berry phases
in Ref. \cite{ACol}.


\emph{III. Hidden SUSY in delta potential problem. } ---A different
structure for hidden SUSY is provided by a one-dimensional delta
potential problem given by
\begin{equation}\label{delta}
    H=-\frac{d^2}{dx^2}-2\beta\delta(x)+\beta^2.
\end{equation}
In the case of the attractive potential with $\beta>0$,
the energy
of the unique  bound state
\begin{equation}\label{psid0}
    \psi_0(x)=\sqrt{\beta}e^{-\beta\vert x\vert}
\end{equation}
is equal to zero due to a special value of the constant term in the
Hamiltonian. For energy values $E>\beta^2$ the wave functions
\begin{equation}\label{WFDD}
    \psi_{k}^{(+)} (x)=
    \left(e^{ikx}+re^{-ikx}\right)\Theta(-x)+te^{ikx}\Theta(x)
\end{equation}
and $\psi_{k}^{(-)}(x)=\psi_k^{(+)}(-x)$ correspond to scattering
states with plane waves incoming, respectively,  from $-\infty$ and
$+\infty$. Here $k=\sqrt{E-\beta ^{2}}>0$, $\Theta(x)$ is a step
function ($\Theta(x)=0$ for $x<0$ and $\Theta(x)=1$ for $x> 0$),
$r(k)=-\beta/(\beta+ik)$ and $ t(k)=ik/(\beta+ik)$ are the
reflection and transmission coefficients. Having a singlet zero
energy bound state and double degeneracy of the energy levels in the
scattering sector, we have  a hidden unbroken $N=2$ SUSY structure.
It can be made explicit by identifying a reflection $R$,
$R\psi(x)=\psi(-x)$, as a grading operator. Unlike the previous
system, here the both supercharges
\begin{equation}\label{QQDD}
    Q_1=-i\left(\frac{d}{dx}+\beta \varepsilon (x)R\right),\qquad
     Q_2=iRQ_1,
\end{equation}
$Q_a=Q_a^\dagger$, $\varepsilon(x)\equiv\Theta(x)-\Theta(-x)$, are
nonlocal due to the presence in them of the reflection operator.
Using the relation $\delta(x)R\psi(x)=\delta(x)\psi(x)$, one finds
that the odd supercharges together with the even Hamiltonian
generate the $N=2$ superalgebra (\ref{Susya}). The bound ground
state (\ref{psid0}) is annihilated by the both supercharges, and is
identified as a bosonic-like state, $R\psi_0(x)=\psi_0(x)$.
Parameterizing the reflection and transmission coefficients as
$r=i\sin\gamma e^{i\gamma}$, $t=\cos\gamma e^{i\gamma}$,
$\sin\gamma=\beta/\sqrt{E}$, $\cos\gamma=k/\sqrt{E}$, one finds that
the scattering information is encoded in the structure of the
distorted plane wave states $ \tilde{\psi}^{(\pm)}_k\equiv
\cos\left(\gamma +k\vert x\vert\right)\pm i\sin kx, $ being the
eigenstates of the supercharge $Q_1$ linear in derivative,
$Q_1\tilde{\psi}^{(\pm)}_k=\pm\sqrt{E}\,\tilde{\psi}^{(\pm)}_k$.

The substitution $\beta\rightarrow -\beta$ yields the case of a
repulsive potential, for which the analog of (\ref{psid0}) is not
normalizable.  This case  is characterized by the hidden broken
$N=2$ SUSY with a typical double degeneracy of the continuous
spectrum with $E>\beta^2$. In the limit $\beta\rightarrow 0$ the
scattering states $\psi^{(+)}_k(x)$ and $\psi^{(-)}_k(x)$ are
transformed into the plane wave solutions with $E>0$, while
$\psi_0(x)/\sqrt{\beta}$ is reduced to a constant wave function
corresponding to a singlet state with $E=0$, and the hidden SUSY of
the delta potential problem is transformed into the hidden unbroken
$N=2$ SUSY of a free particle on a line.


\emph{IV. Hidden nonlinear SUSY in PT system.}
---Now consider the quantum PT potential problem \cite{PT,Ros} given by the
Hamiltonian
\begin{equation}\label{PT}
    H_\lambda=-\frac{d^2}{dx^2}-\lambda(\lambda+1)\frac{\omega^2}{\cosh^2\omega
    x}+\lambda^2\omega^2
\end{equation}
with two real parameters $\lambda$ and $\omega$. It factorizes, $
    H_\lambda=-{\cal D}_{-\lambda}{\cal D}_\lambda,
$ via the first order differential operators,
\begin{equation}\label{Dlam}
    {\cal D}_\lambda=\frac{d}{dx}+\lambda\omega\tanh \omega x,\qquad
    {\cal D}_{-\lambda}=-{\cal D}_\lambda^\dagger,
\end{equation}
satisfying the relations
\begin{equation}\label{DDH}
    {\cal D}_{-\lambda}{\cal D}_{\lambda}={\cal D}_{\lambda+1}{\cal
    D}_{-(\lambda+1)}+(2\lambda+1)\omega^2,
\end{equation}
\begin{equation}\label{Da0}
    {\cal D}_{\lambda}=C^{-1}{\cal D}_{\lambda-1}C=
    C^{-\lambda}\frac{d}{dx}C^\lambda ,
\end{equation}
$C\equiv\cosh\omega x$. At integer $\lambda$ the second equality
from Eq. (\ref{Da0}) is a consequence of the iterative application
of the first equality. It is this property that is behind the
reflectionless  nature of the PT potential with $\lambda=l$, $l\in
\Z$, due to which system (\ref{PT}) turns out to be related to the
nonlinear integrable systems \cite{CKS}. We put $\lambda=l$
assuming that $l=1,2,\ldots$.

System (\ref{PT}) with $\lambda=l$ possesses $l$ bound states.  Its
continuous spectrum is doubly degenerate except the first level with
$E_{l}(0)=l^2\omega^2$, see below.  This indicates on existence in
it of a hidden nonlinear SUSY \cite{ParaP,AIS}. Again, we identify
the reflection, $R$, $[H_l,R]=0$, as a grading operator. The system
has a local integral of motion
\begin{equation}\label{Pn}
    {\cal P}_l=i{\cal D}_{-l}{\cal D}_{-l+1}\ldots {\cal D}_l,
\end{equation}
${\cal P}_l={\cal P}_l^{\dagger}$, being the order $(2l+1)$
differential operator \cite{comment}.
 It satisfies the relation ${\cal P}_l^2=P_{2l+1}(H_l)$,
where the order $2l+1$ polynomial is
\begin{equation}\label{PHn}
    P_{2l+1}(H_l)=(H_l-l^2\omega^2)\prod_{n=0}^{l-1}
    \left(H_l-(l^2-(l-n)^2)\omega^2\right)^2.
\end{equation}
Since $R$ anticommutes with integral (\ref{Pn}), we identify the
operators
\begin{equation}\label{QQPT}
     Q_{l,1}={\cal P}_l,\qquad
    Q_{l,2}=iRQ_{l,1},
\end{equation}
$Q_{l,a}=Q_{l,a}^\dagger$,  as the supercharges. Together with the
Hamiltonian they generate a nonlinear (polynomial) superalgebra of
the order $2l+1$,
\begin{equation}\label{susynon}
    \{Q_{l,a},Q_{l,b}\}=2\delta_{ab}\, P_{2l+1}(H_l),\qquad
    [H_l,Q_{l,a}]=0,
\end{equation}
whose form  coincides with that of the hidden SUSY of the order
$2(l+1)$ parabosonic oscillator \cite{ParaP}.

The (non-normalized) bound states have the form
$\psi_{l,0}=\cosh^{-l} \omega x$ and
\begin{equation}\label{PTbs}
\psi_{l,n}(x)={\cal D}_{-l}{\cal D}_{-l+1}\ldots
    {\cal D}_{-l+n-1}\cosh^{n-l}\omega x,
\end{equation}
where $n=1,\ldots, l-1$. These are the singlet states corresponding
to the $n$ discrete energy levels
\begin{equation}\label{EnPT}
    E_{l,n}=(l^2-(l-n)^2)\omega^2,\qquad n=0,\ldots, l-1.
\end{equation}
They are annihilated by the supercharges $Q_{l,a}$, and have a
definite parity, $R\psi_{l,n}=(-1)^n\psi_{l,n}$.

In correspondence with reflectionless nature of the system, the
functions
\begin{equation}\label{PTcont}
    \psi _{l,k}^{(\pm)}(x)=\mathcal{D}_{-l}
    {\cal D}_{-l+1}\ldots \mathcal{D}_{-1}
    \exp (\pm ikx)
\end{equation}
with $k\geq 0$ describe the scattering states of the energy
$E_l(k)=k^2+l^2\omega^2$, and satisfy the relation
$R\psi^{(\pm)}_{l,k}=(-1)^l\psi^{(\mp)}_{l,k}$. They are the
eigenstates of the local supercharge $Q_{l,1}$,
\begin{equation}\label{QPTcon}
    Q_{l,1}\psi _{l,k}^{(\pm)}=\mp(-1)^{l}kE_1(k) E_2(k)\ldots
    E_l(k)\psi _{l,k}^{(\pm)},
\end{equation}
$E_n(k)\equiv k^2+n^2\omega^2,$ $n=1,\ldots, l$. The states $\psi
_{l,k}^{(+)}$ and $\psi _{l,k}^{(-)}$ with $k>0$ form a SUSY
doublet.  The SUSY singlet state $\psi^l_0\equiv\psi_{l,0}^{(+)}=
\psi _{l,0}^{(-)}$ is annihilated by the both supercharges. Eqs.
(\ref{PTbs}) and (\ref{EnPT})  at $n=l$ reproduce this singlet
state, being a polynomial of order $l$ in $\tanh \omega x$, and its
corresponding energy.  Note that the simplest (``fermionic") singlet
state $\psi^1_0(x)=\tanh\omega x$ appears as a kink solution in the
2-dimensional  $\varphi^4$ field theory, for which the Schr\"odinger
equation with PT potential plays a role of a stability equation
\cite{Jack}.

 The scattering data can
be extracted from the eigenstates of the supercharge $Q_1$. Taking
the limits $x\rightarrow \mp \infty$ in (\ref{PTcont}), one gets
$\psi^{(+)}_{l,k}(x) \rightarrow A^{\mp}_{l}(k)e^{ikx}$,
\begin{equation}\label{ApmPT}
    A^\mp_{l}(k)=(ik\pm l\omega)(ik\pm (l-1)\omega)\ldots (ik\pm
    \omega).
\end{equation}
As a result we find the transmission coefficient
$t_l(k)=A^+_l(k)/A^-_l(k)$.  It can be presented in the form $
t_l(k)=\exp\left(-2i(\delta_{1,k}+\ldots \delta_{l,k})\right),$
$e^{-i\delta_{n,k}}=(n\omega -ik)/\sqrt{E_{n,k}}$.

Thus, the PT system with  parameter $\lambda=l$ has the hidden
unbroken $N=2$ nonlinear SUSY characterized by the polynomial
superalgebra (\ref{susynon}), and by the $l+1$ singlet SUSY states
of alternating parity.

Though the structure of the hidden SUSY in the PT and the delta
potential problems is essentially different, the Hamiltonian of
the the latter system can be obtained via a limit procedure from
the former system of a generic form with $\lambda\in \R$.
Rescaling the coordinate variable, $x\rightarrow
\omega\lambda\beta^{-1}x$, and taking the double limit
$\lambda\rightarrow 0$, $\omega\rightarrow \infty$,
$\lambda\omega=\beta$, we reduce Hamiltonian (\ref{PT}) to
(\ref{delta}).


\emph{V. Summary and discussion.} ---Let us summarize and discuss
the obtained results. The revealed hidden SUSY of the three
quantum {\it bosonic} systems with {\it local} Hamiltonians is
based on the existence of the grading operator, which is a {\it
nonlocal} integral of motion being a reflection or a twisted
reflection operator, and of a nontrivial integral anticommuting
with it. In the bound state AB and delta potential problems the
supercharges have a nature of a square root of the Hamiltonian,
while in the PT system with parameter $\lambda=l$ they are of the
square root of the order $2l+1$ polynomial in $H_l$ nature. In the
AB and PT systems one of the self-adjoint supercharges is {\it
local}, but in the delta potential problem the both self-adjoint
supercharges are {\it nonlocal} operators.

In the bound state AB effect, Eq. (\ref{AB-0}) being applied to the
reflection operator $R$ yields a well defined twisted reflection
operator only for integer and half-integer values of the parameter
$\alpha$. Exactly in these two special cases the system enjoys the
hidden unbroken ($\alpha=n$), or broken ($\alpha=n+\frac{1}{2}$)
$N=2$ SUSY characterized by the minimum, $E=0$, and maximum,
$E=\frac{1}{4}$, energy values of the ground state. A hidden SUSY is
also present in the AB system extended to the cylindrical geometry
and given  by the Hamiltonian $H_{cyl,\alpha}=H_\alpha + p_z^2 $,
where $p_z=-i\frac{d}{dz}$, $-\infty<z<\infty$.  For such a system
the hidden $N=2$ SUSY exists, again, only for $2\alpha\in\Z$. It is
characterized by the grading operator $\Gamma=R_zR_\alpha$,
$\Gamma\psi(z,\varphi)=e^{-2i\alpha\varphi}\psi(-z,-\varphi)$. The
supercharges are the nonlocal operators $Q_{\alpha,1}=(p_\varphi
+\alpha) + i\Gamma p_z$, $Q_{\alpha,2}=i\Gamma Q_{\alpha,1}$,
$Q_{\alpha,a}=Q_{\alpha,a}^\dagger$, generating the superalgebra of
the form (\ref{Susya}). In the case of the planar AB system with
$\alpha=n$, a hidden $N=2$ SUSY can be revealed following the same
line. However, for half-integer values of $\alpha$ the action of the
supercharge operators takes off some of the energy eigenstates from
the Hamiltonian  domain. Due to these complications, the analysis of
the hidden SUSY in the planar AB effect requires a special treatment
related to the problem of a self-adjoint extension \cite{ABsusy}.

The hyperbolic PT potential can be treated as a limit case of some
doubly periodic elliptic functions with a real period tending to
infinity \cite{Ellip,AGI}. The quantum problems with periodic
potential given by an elliptic function appear, in particular, in
solid state physics as the crystals models. It would be
interesting to look for a hidden supersymmetry in such a class of
periodic quantum mechanical systems \cite{Note1}. They are also of
interest for supersymmetric quantum mechanics \cite{SUSYper},
Kaluza-Klein \cite{KK}, and preheating after inflation theories
\cite{infl}.

Since the three considered related bosonic systems reveal a hidden
$N=2$  SUSY of linear or nonlinear  form, it has also to be present
as a hidden, additional SUSY in the corresponding ordinary
supersymmetric systems obtained via the extension by fermionic
(spin) degrees of freedom.  Unlike the fictitious double SUSY of
superconformal mechanics model observed in Ref. \cite{SUSYconf}, the
supersymmetric extentions of these three systems have to possess a
true double supersymmetry.

It seems likely that the hidden SUSY should also reveal itself in
nonlinear integrable systems and field theoretical models. The
natural candidates for investigation in such a direction are the
systems to which the PT quantum potential problem is related,
namely, the Korteweg-de Vries  equation \cite{Ros}, and  the
2-dimensional $\varphi^4$ and sine-Gordon  field theories
\cite{Jack}.
 Another class of the systems
corresponds to the (2+1)D models with Chern-Simons term inducing a
spin-statistics transmutation based on the AB effect
\cite{Anyons,Note2}.

\vskip 0.1cm \emph{Acknowledgements.} MP thanks J. Zanelli and J.
Gomis for useful discussions, and CECS, where a part of this work
was done, for hospitality extended to him. The work of FC was
supported in part by CONICYT PhD Program Fellowship, and that of MP
by FONDECYT-Chile (project 1050001).


\end{document}